# Enterprise Data Modelling Methodologies:
## A Comparative Analysis of Inmon, Kimball, and Data Vault

Issar Arab
Adrem Data Lab, Department of Computer Science
University of Antwerp
Issar.arab@tum.de

## ABSTRACT

The design and governance of enterprise data warehouses constitute foundational decisions in modern data-driven organisations, with long-term impact for analytical capability, operational agility, and regulatory compliance. This paper presents a structured comparative analysis of three prevailing data warehousing methodologies: the Inmon approach, the Kimball approach, and Data Vault. The paper first establishes the technical foundations common to all three enterprise frameworks, in particular the distinction between Online Transaction Processing (OLTP) and Online Analytical Processing (OLAP) systems, the principles of relational normalisation, and the core techniques of entity-relationship and dimensional data modelling. The comparative analysis examines each methodology across a set of dimensions including architectural philosophy, modelling technique, scalability, agility, query performance, audit capability, and suitability for different organisational profiles. Findings indicate that no single methodology is universally optimal; rather, the appropriate choice is contingent on an organisation's scale, regulatory environment, analytical maturity, and tolerance for upfront architectural investment. This paper concludes with a synthesis of decision criteria to guide practitioners and researchers in selecting the methodology most aligned with their strategic objectives.

***Keywords:*** *data warehousing, enterprise data modelling, Inmon methodology, Kimball dimensional modelling, Data Vault, OLAP, OLTP, data architecture*

# 1. INTRODUCTION

The exponential growth of organisational data, driven by the proliferation of transactional systems, digital channels, and interconnected operational platforms, has elevated data warehousing from a technical infrastructure concern to a strategic imperative. Organisations increasingly recognise that the capacity to integrate, govern, and analyse data at enterprise scale directly conditions their competitive positioning, operational efficiency, and regulatory compliance. At the heart of this capability lies the data warehouse: a structured repository designed to consolidate heterogeneous data sources into a unified analytical environment (Inmon, 2005).

Yet the design of such foundational infrastructure is far from straightforward. Practitioners and architects must navigate a complex landscape of competing methodological frameworks, each embodying distinct assumptions about how data should be structured, integrated, and made accessible. Three methodologies have emerged as dominant paradigms in both academic discourse and industry practice: the Inmon approach, which advocates a centralised, top-down enterprise data warehouse; the Kimball approach, which champions a bottom-up, business-driven architecture of conformed dimensional data marts; and Data Vault, a hybrid methodology engineered for scalability, agility, and auditability in dynamic data environments (Linstedt & Olschimke, 2015).

Each of these methodologies reflects a coherent but divergent theory of enterprise data management. The Inmon paradigm prioritises definitional authority and long-term governance, accepting a higher upfront investment in exchange for a single, normalised source of truth (Inmon, 2005). The Kimball paradigm places business usability and time-to-value at the forefront, organising data around business processes and delivering analytical capability incrementally (Kimball & Ross, 2013). Data Vault, developed by Dan Linstedt and evolved over two decades of industry practice, separates business identity, relational context, and descriptive attributes into distinct structural components—enabling the architecture to absorb change without requiring schema redesign (Linstedt & Olschimke, 2015).

The strategic importance of enterprise data infrastructure is further reinforced by its role as the foundation for advanced predictive analytics, a domain in which the quality and governance of underlying data architectures directly conditions both system and model performance (Ravi et al. 2024; Arab et al., 2025; Arab et al., 2019).

The selection of an appropriate methodology is consequential. Organisations that adopt a framework misaligned with their scale, regulatory environment, or analytical maturity risk incurring significant technical debt, prolonged delivery timelines, or governance failures. Despite the practical importance of this decision, comparative analyses that rigorously evaluate all three methodologies across a unified set of dimensions remain relatively limited in the academic literature. Much of the existing reports are either vendor-oriented, limited to pairwise comparisons, or insufficiently grounded in the technical foundations of data modelling.

This paper addresses that gap by providing a comprehensive, academically rigorous comparative analysis of the Inmon, Kimball, and Data Vault methodologies. The analysis is structured as follows: Section 2 establishes the technical foundations—OLTP versus OLAP processing paradigms, relational normalisation theory, and the principal data modelling techniques—upon which all three methodologies build. Section 3 examines each methodology in depth, analysing architectural design, structural components, advantages, limitations, and illustrative applications. Section 4 presents a consolidated benchmark comparison across twelve evaluation criteria. Section 5 offers concluding remarks and directions for future research.

# 2. TECHNICAL FOUNDATIONS OF DATA MODELLING

Before examining enterprise data warehousing methodologies, it is necessary to establish the technical substrate on which they are constructed. This section introduces the processing paradigms, design principles, and modelling techniques that inform the architectural choices examined later in the paper.

## 2.1 The Data Warehouse

A data warehouse is a centralised, integrated, time-variant, and non-volatile repository designed to consolidate data from multiple heterogeneous operational sources for analytical and decision-support purposes (Inmon, 2005). Unlike operational databases, a data warehouse is structurally optimised for complex analytical queries rather than transactional throughput. Data is typically ingested through Extract, Transform, and Load (ETL) pipelines that cleanse, conform, and integrate data from disparate source systems before persisting it in a consistent and structured form.

Modern data warehouses typically adopt layered architectures that separate data ingestion, integration storage, and presentation layers. This architectural separation enhances scalability, query performance, and governance by assigning distinct responsibilities to each layer and enabling independent optimisation (Kimball & Ross, 2013). The functional benefits of a well-architected data warehouse include: the integration of diverse data sources into a unified analytical repository; support for comprehensive historical analysis spanning extended time horizons; improved data consistency and quality through standardised transformation rules; and the isolation of analytical workloads from transactional systems, preserving operational system performance.

## 2.2 Online Transaction Processing versus Online Analytical Processing

In contemporary data ecosystems, transactional and analytical workloads are handled by two structurally distinct processing paradigms: Online Transaction Processing (OLTP) and Online Analytical Processing (OLAP). Each is optimised for fundamentally different operational objectives, and the distinction between them is a prerequisite for understanding the separation of concerns that underpins all three methodologies examined in this paper.

OLTP systems are engineered to support real-time operational processes. They manage high-frequency transactions such as sales entries, inventory adjustments, customer account modifications, and payment authorisations. The defining requirements of OLTP environments are high concurrency, low-latency write operations, and strict data integrity. To satisfy these requirements, OLTP databases are typically normalised to the third normal form (3NF), minimising data redundancy and ensuring transactional consistency (Codd, 1970).

OLAP systems, on the other hand, are optimised for analytical workloads. They support complex queries executed over large volumes of historical and aggregated data to facilitate strategic reporting, trend analysis, forecasting, and executive decision-making. OLAP environments prioritise read performance and query expressiveness over transactional throughput. Accordingly, they commonly employ denormalised or multidimensional data structures, such as star and snowflake schemas, that reduce join complexity and accelerate aggregation operations (Chaudhuri & Dayal, 1997). Table 1 provides a structured comparison of the two paradigms across key dimensions.

| Feature | OLTP | OLAP |
|---|---|---|
| **Purpose** | Supports real-time operational transactions | Supports analytical queries and decision-making |
| **Data Type** | Current, granular transactional data | Historical, aggregated data |
| **Query Type** | Short, simple, high-frequency queries | Complex, long-running analytical queries |
| **Operations** | INSERT, UPDATE, DELETE | SELECT, GROUP BY, AGGREGATE |
| **DB Design** | Highly normalised (3NF) | Dimensional or denormalised models |
| **Performance Goal** | Fast transaction processing | Fast analytical query performance |
| **Users** | Operational staff and applications | Analysts, managers, BI tools |
| **Example** | Point-of-sale systems, CRM, e-commerce platforms | Data warehouses, dashboards, reporting systems |

*Table 1. Comparative Summary of OLTP and OLAP Processing Paradigms*

## 2.3 Relational Normalisation

Normalisation is a systematic database design technique for structuring relational schemas to reduce redundancy and eliminate data anomalies. Its core theoretical foundation is based on the concept of functional dependency (FD), which formalises the semantic relationships between attributes in a relation. A functional dependency $\alpha \rightarrow \beta$ denotes that for any group of tuples in a relation R, agreement on the attributes in $\alpha$ implies agreement on the attributes in $\beta$—that is, $\alpha$ functionally determines $\beta$ (Codd, 1972). Functional dependencies are the analytical instruments through which normalisation identifies and eliminates structural anomalies.

The practical motivation for normalisation is the prevention of three classes of anomalies that afflict poorly structured schemas. Update anomalies arise when the same fact is recorded redundantly, requiring multiple coordinated modifications. Insertion anomalies prevent the recording of new facts unless unrelated data is also present. Deletion anomalies cause the unintended loss of information when records are removed.

Functional dependencies guide the decomposition process in normalization. When breaking a relation into smaller relations, we rely on FDs to identify redundancy and dependency patterns between attributes. However, decomposition must satisfy two important properties. First, it should be lossless, meaning that no information is lost when the decomposed relations are joined back together. Second, it should ideally be dependency-preserving, meaning that all functional dependencies from the original relation can still be enforced in the decomposed schema without requiring additional joins.

The normalisation process proceeds through a hierarchy of normal forms. First Normal Form (1NF) requires that all attribute domains contain only atomic values and that no repeating groups exist. Second Normal Form (2NF) eliminates partial dependencies, ensuring that every non-key attribute is fully functionally dependent on the entire primary key rather than a subset of it. Third Normal Form (3NF) eliminates transitive dependencies, guaranteeing that non-key attributes depend exclusively on the primary key and not on other non-key attributes (Date, 2003). In practice, 3NF represents the standard target for OLTP schemas, balancing integrity and update efficiency. The Inmon methodology applies 3NF to the enterprise data warehouse layer, while Data Vault employs its own form of extreme normalisation through the separation of business keys, relationships, and descriptive context.

## 2.4 Data Modelling Approaches

Data modelling is the disciplined process of defining and structuring data elements, their attributes, and their interrelationships within an information system. A data model translates business semantics into a logical and physical blueprint that guides database design and ensures alignment between organisational requirements and technical implementation (Simsion, 2007). Two modelling paradigms are of particular relevance to the methodologies examined in this paper: entity-relationship modelling and dimensional modelling.

### *2.4.1 Entity–Relationship Modelling*

Entity-Relationship (ER) modelling, introduced by Chen (1976), is a conceptual technique for representing the logical structure of relational databases. It decomposes a domain into entities, meaning distinct real-world objects or concepts (e.g. customer, product, order), their attributes (e.g. customer_id, product_name), and the relationships that connect them (e.g. a customer places an order). ER models provide a semantically rich, normalisation-compatible representation of operational domains and constitute the standard modelling technique for OLTP systems. The Inmon methodology applies ER modelling principles to the construction of the enterprise data warehouse, employing 3NF schemas that closely resemble the structure of operational source systems.

A closely related notation is the Unified Modelling Language (UML), a standardised modelling language widely adopted during the design phase of the software development lifecycle to represent object-oriented systems, database structures, and application architectures (Arab et al., 2018; Arab & Bourhnane, 2018). UML class diagrams, in particular, serve as a formal mechanism for modelling entity relationships and are broadly applicable to the conceptual design of data-intensive systems. Despite its widespread adoption, UML has been criticized for its lack of formal semantics, a limitation that has motivated several proposals for formalisation in the research community (Falah et al., 2017). Notably, the relationship between UML and legacy design methodologies, such as Merise, offers a further bridge between conceptual modelling and relational database design, particularly in enterprise contexts where both paradigms coexist (Arab et al., 2018).

### *2.4.2 Dimensional Data Modelling*

Dimensional modelling, popularised by Ralph Kimball, is a technique specifically designed to optimise analytical querying in data warehouse environments. In contrast to the normalised structures favoured in operational systems, dimensional models organise data into two principal components: fact tables, which store quantitative, measurable business metrics (e.g. revenue, units sold) along with foreign keys referencing dimension tables; and dimension tables, which provide the descriptive context (e.g. customer name, product category, date) that enables filtering, grouping, and slicing of the quantitative facts (Kimball & Ross, 2013).

Controlled denormalisation is a deliberate design choice in dimensional modelling: it reduces join complexity, improves query performance, and produces schemas that are intuitively navigable by business analysts.

Two schema patterns are commonly employed in dimensional modelling. The star schema (Figure 1) places a central fact table in direct relationship with a set of fully denormalised dimension tables, minimising join depth and maximising query performance. The snowflake schema (Figure 2) extends this structure by normalising dimension tables into hierarchical sub-dimensions, reducing storage redundancy at the cost of increased join complexity. Both patterns are widely used in OLAP environments and are central to the Kimball methodology.

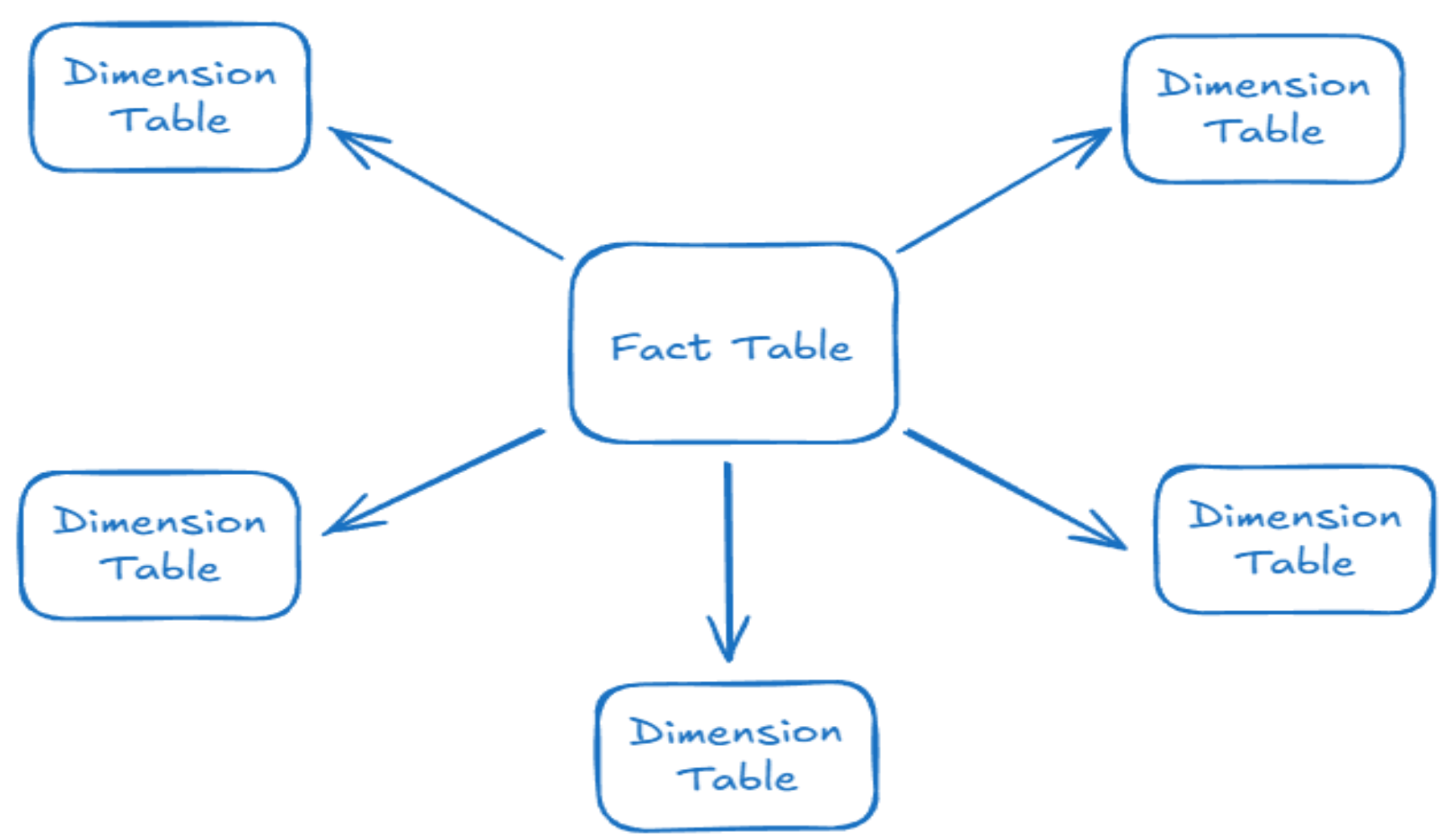

*Figure 1. Star Schema — Fact table at centre with denormalised dimension tables*

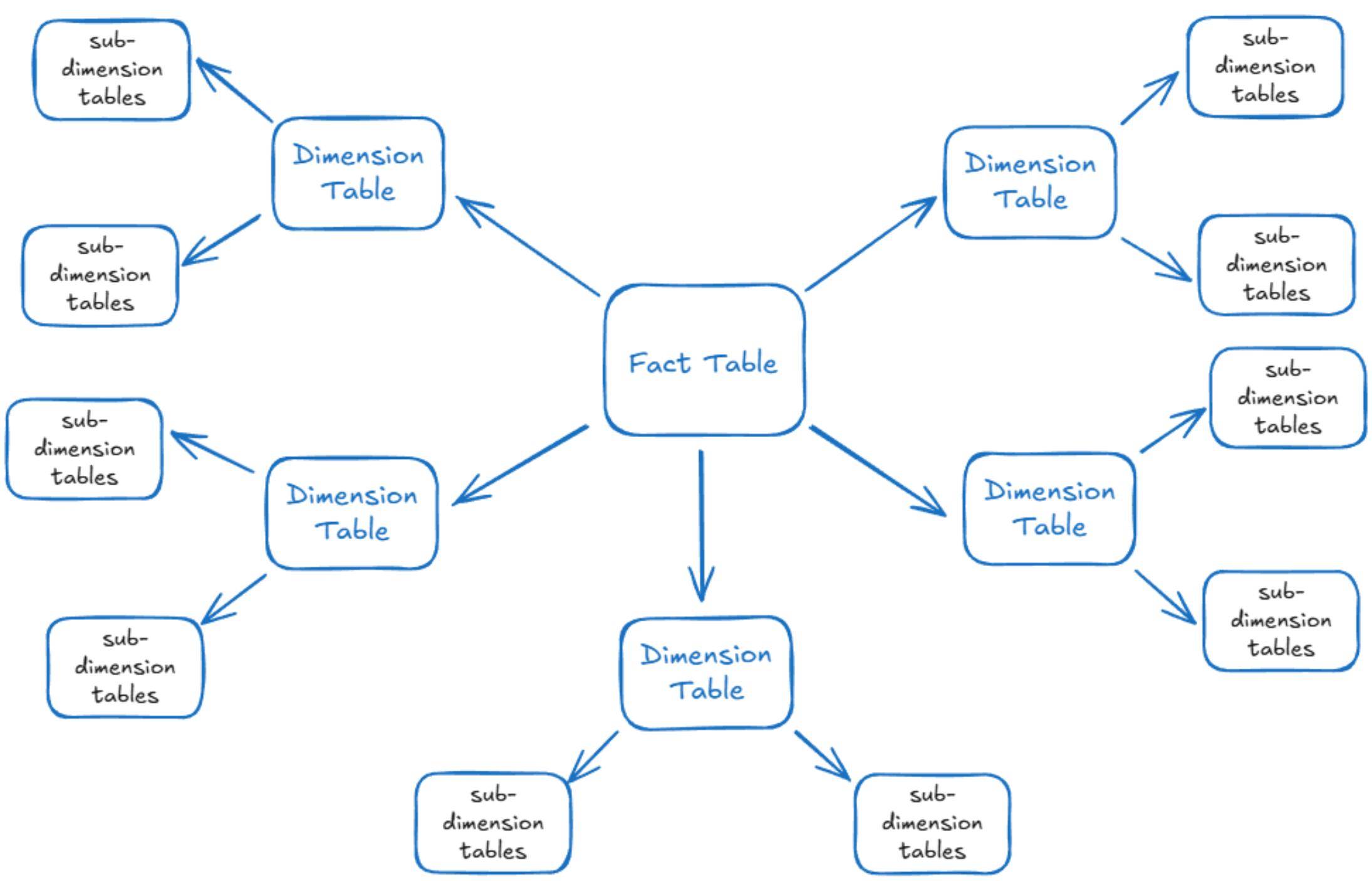

*Figure 2. Snowflake Schema — Normalised dimension hierarchies extending from the fact table*

# 3. ENTERPRISE DATA WAREHOUSING METHODOLOGIES

Building upon the technical foundations established in the preceding section, this section examines three dominant enterprise data warehousing methodologies in depth. Each methodology is analysed with respect to its foundational philosophy, architectural design, structural components, advantages, limitations, and illustrative application. While these methodologies share the common objective of enabling enterprise-scale analytical capability, they differ fundamentally in how they approach data integration, governance, and the sequencing of architectural decisions.

## 3.1 The Inmon Approach

### *3.1.1 Overview and Design Philosophy*

The Inmon approach, originally articulated by Bill Inmon and widely regarded as the classical enterprise data warehousing framework, is characterised as a top-down methodology. Its central premise is that the first step in enterprise data warehousing is the construction of a centralised Enterprise Data Warehouse (EDW) modelled in the third normal form. This repository integrates data from all organisational source systems into a single, internally consistent store that functions as the authoritative source of truth for the enterprise. Only after this normalised foundation has been established, subject-oriented data marts are then derived to support the analytical and reporting needs of individual business units (Inmon, 2005).

This sequencing reflects the central principle of the methodology: analytical systems should not be built independently but should instead be derived from a unified integration layer. In this approach, the data warehouse serves as that integration layer. According to Bill Inmon, a data warehouse is defined as subject-oriented, integrated, time-variant, and non-volatile. These four characteristics distinguish a data warehouse from operational transaction systems. The normalized EDW is structured specifically to ensure that these characteristics are maintained. As a result, the approach prioritizes consistent definitions, comprehensive historical data, and strong enterprise governance, even if this slows the speed at which analytical systems can be delivered.

### *3.1.2 Architectural Design*

The Inmon architecture enforces a clear separation between data integration and data consumption layers. Data flows unidirectionally from operational source systems through a centralised ETL and integration layer into the normalised EDW, and subsequently into subject-specific data marts that serve end-user analytical needs. The ETL and integration layer performs centralised extraction, cleansing, standardisation, and conforming of data across heterogeneous sources. The Enterprise Data Warehouse stores integrated data at an atomic level in a 3NF schema, preserving all historical records through time-variant design and serving as the enterprise's single source of truth. Downstream data marts—organised by subject area such as Sales, Finance, or Operations—are derived from the EDW and adopt dimensional or otherwise query-optimised schemas appropriate to their intended consumer base. End-user reporting tools and dashboards interface exclusively with the mart layer, not with the EDW directly.

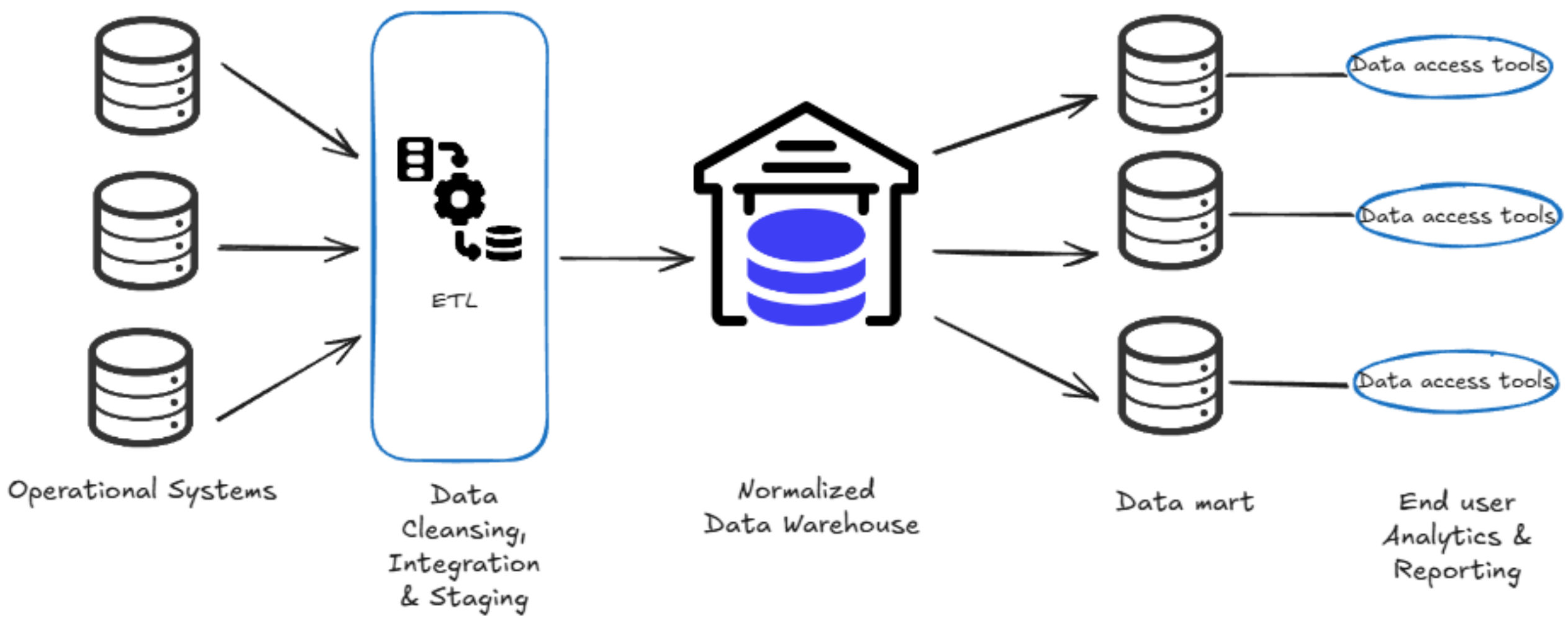

*Figure 3. Inmon Architecture, a normalised EDW feeding downstream subject-oriented data marts*

### *3.1.3 Advantages and Limitations*

The foremost strength of the Inmon approach lies in its capacity to enforce enterprise-wide consistency. By mandating that all analytical environments derive from a single, normalised EDW, the methodology eliminates the definitional discrepancies that commonly arise when business units independently construct siloed data environments. This architectural rigour is particularly valuable in regulated industries, such as financial services, insurance, and healthcare, where data lineage, auditability, and metric consistency are subject to formal compliance requirements. Table 2 summarises the principal advantages and limitations of the approach.

| Advantages | Limitations |
|---|---|
| • Enterprise-wide consistency: unified definitions, metrics, and reporting standards enforced across all organisational units.<br>• Strong data governance: centralised control enables rigorous enforcement of data standards, quality, and lineage.<br>• High data integrity: 3NF design minimises redundancy, reduces update anomalies, and preserves referential consistency.<br>• Scalability for complex enterprises: well-suited to large, multi-departmental, or regulated organisations.<br>• Reliable historical analysis: time-variant, non-volatile data supports accurate trend analysis and audit trails. | • Extended time-to-value: significant upfront modelling and integration effort is required before analytical capability is delivered.<br>• Higher initial investment: costs are front-loaded in the architectural and design phases.<br>• Reduced agility: structural schema changes are slow and costly in dynamic business environments.<br>• Suboptimal for direct querying: analytical performance depends on the downstream derivation of purpose-built data marts. |

*Table 2. Advantages and Limitations of the Inmon Approach*

The methodology's principal limitation is its extended time-to-value. Because the full EDW must be modelled, integrated, and validated before downstream data marts can be delivered, the approach demands significant upfront investment in both financial resources and specialised modelling expertise. This front-loading renders the methodology less suitable for organisations operating in dynamic environments where rapid analytical capability is a competitive necessity. Furthermore, the rigidity of the normalised schema can impede responsiveness to structural changes in source systems or evolving analytical requirements.

#### *3.1.4 Illustrative Application*

To contextualise the Inmon approach within a practical setting, consider a large-scale e-commerce organisation whose operations generate data across multiple transactional domains—including order management, inventory control, and digital marketing platforms. Under this methodology, data from these heterogeneous operational systems is consolidated into a centralised EDW prior to any analytical use. Within the EDW, data is stored in third normal form to ensure cross-domain consistency, minimise redundancy, and produce a unified canonical representation of core business entities. Once the EDW is established as a stable and trusted foundation, the organisation derives subject-specific data marts—for Sales, Marketing, and Procurement; for example, each adopting a dimensional schema optimised for the analytical patterns of its intended consumer base. Every metric or dimension presented in a business report is ultimately traceable to the enterprise's single source of truth.

## 3.2 The Kimball Approach

### *3.2.1 Overview and Design Philosophy*

The Kimball approach, developed by Ralph Kimball in the 1990s and formalised in successive editions of The Data Warehouse Toolkit (Kimball & Ross, 2013), represents a bottom-up methodology for enterprise data warehousing centred on dimensional modelling. In contrast to approaches that prioritise the prior construction of a centralised enterprise data warehouse, the Kimball methodology focuses on delivering analytical capability incrementally through the development of business-oriented data marts.

Under this approach, development begins with dimensional data marts aligned to specific organisational business processes, such as sales transactions, inventory management, or marketing activities. Each mart is designed to support a clearly defined analytical domain. Over time, these marts are integrated into a coherent enterprise data warehouse through the disciplined use of conformed dimensions—standardised dimensional structures that are shared across multiple marts. Conformed dimensions enable consistent definitions for core analytical entities such as customer, product, and time, thereby allowing analytical queries to span multiple business processes (Kimball & Ross, 2013).

The central principle of the Kimball methodology is that the enterprise data warehouse should be understood not as a single, pre-constructed repository but as the integrated collection of all conformed dimensional data marts. This perspective places strong emphasis on business usability and rapid analytical delivery. Consequently, the methodology promotes close collaboration with business stakeholders, iterative development cycles, and the use of schema structures that are readily understandable by non-technical analysts.

### *3.2.2 Architectural Design*

The Kimball methodology is implemented through a bus architecture, in which integration across the data warehouse is achieved through shared conformed dimensions rather than through a centralised normalised repository. In this architecture, data flows from operational source systems into a staging environment, where it undergoes extraction, transformation, and loading before being distributed to subject-specific dimensional data marts (Figure 4). Each mart is organised around a star/snowflake schema designed to support a particular business process.

Conformed dimensions are shared across multiple marts and function as the primary integration mechanism of the architecture. By maintaining consistent dimensional definitions across the analytical environment, the bus architecture enables analysts to perform cross-functional queries spanning multiple business processes without requiring a separate enterprise integration layer. Business intelligence tools, dashboards, and analytical applications therefore query the dimensional mart layer directly, benefiting from structures that are optimised for read-intensive analytical workloads.

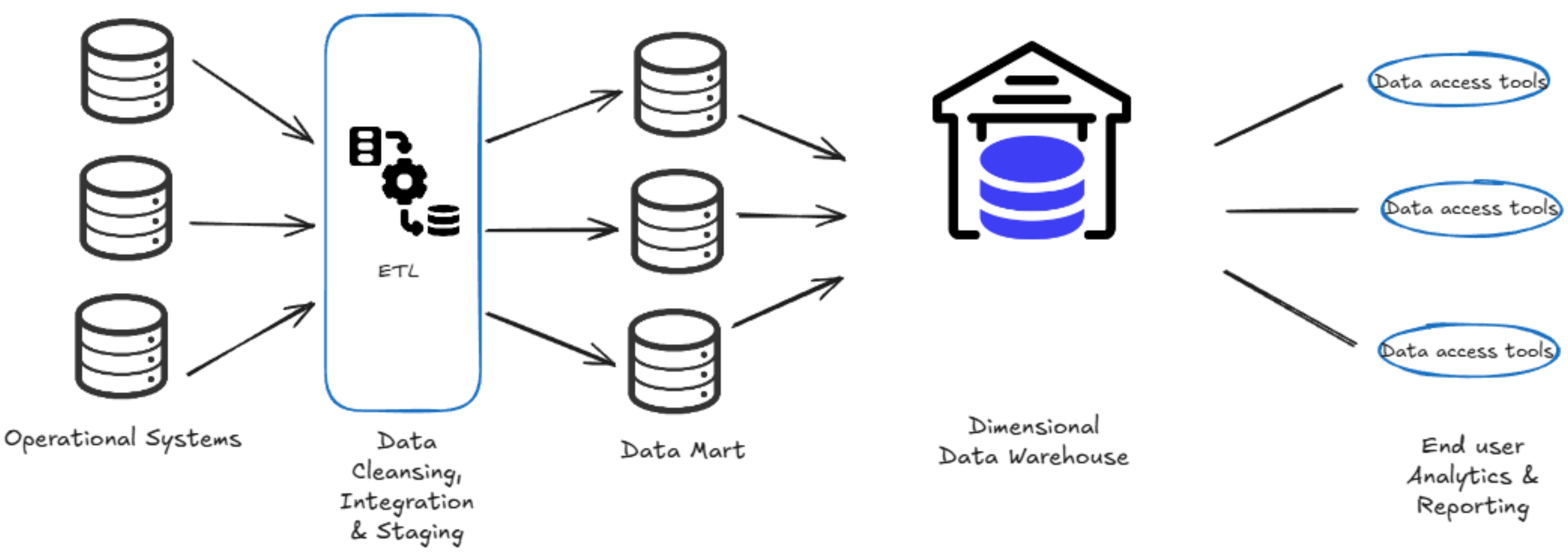

*Figure 4. Kimball's approach – creating data marts first and then developing a data warehouse incrementally from independent data marts.*

This architecture enables rapid and incremental deployment of analytical capabilities. However, it also requires strict governance of dimensional standards as the number of marts increases. Without careful coordination of conformed dimensions, the architecture may gradually fragment into a collection of siloed analytical databases with inconsistent definitions. A failure mode sometimes described in the literature as a 'data swamp' (Inmon, 2016).

### *3.2.3 Advantages and Limitations*

A primary advantage of the Kimball approach is its ability to deliver analytical capability quickly. By focusing initially on a single, clearly defined business process, a dimensional data mart can often be implemented within weeks of project initiation. This incremental delivery model contrasts with methodologies that require the construction of a comprehensive enterprise data warehouse before analytical systems can be deployed. The use of star schemas further accelerates analytical adoption by providing data structures that are intuitive for business analysts to navigate. Because dimensional schemas align closely with common business concepts and reporting patterns, they can be readily accessed through business intelligence tools without requiring advanced database expertise. Table 3 presents detailed advantages and limitations of the approach.

| Advantages | Limitations |
|---|---|
| • Fast time-to-value through incremental, subject-oriented data mart delivery.<br>• High query performance: star schema design minimises join complexity for OLAP workloads.<br>• Business-friendly models: intuitive fact-dimension structures accessible to non-technical analysts.<br>• Lower upfront investment relative to a full enterprise EDW initiative.<br>• Strong alignment with discrete, well-defined business processes and reporting requirements. | • Data redundancy inherent in denormalised star schema structures across marts.<br>• Complex ETL transformation logic required to reshape normalised OLTP data into dimensional models.<br>• Governance challenges arise as the number of marts scales without strict dimensional standards.<br>• Limited flexibility when source system structures change frequently, requiring schema refactoring.<br>• No native raw-data audit layer; complete historical lineage requires explicit SCD implementation. |

*Table 3. Advantages and Limitations of the Kimball Approach*

The methodology's most significant limitation is its sensitivity to change. Because data marts are purpose-built for specific user requirements and business processes, structural changes in source systems or the introduction of new analytical questions can necessitate substantial ETL refactoring and dimensional schema redesign. At enterprise scale, maintaining consistency across a large number of independently developed marts requires considerable governance investment; the failure to enforce conformed dimension standards can erode the integrative value of the Kimball bus architecture.

### *3.2.4 Illustrative Application*

Consider an online retail organisation seeking rapid insight into its sales performance. Rather than investing in an enterprise-wide normalised repository, the organisation initiates a Sales Data Mart using a star schema. The central fact table, fact_sales, stores quantitative measures including sales amount, quantity, discount, and net profit margin. Surrounding dimension tables—date, customer, product, store, and channel—provide the descriptive context necessary for filtering, grouping, and slicing these metrics. This mart can be delivered rapidly, enabling immediate reporting on monthly revenue, regional performance, and customer purchasing patterns. As analytical demand expands, additional marts for marketing campaign analysis and inventory management are built and integrated through conformed Customer, Date, and Product dimensions, progressively constructing an enterprise-wide analytical environment without requiring upfront centralised design.

## 3.3 Data Vault

### *3.3.1 Overview and Design Philosophy*

Data Vault, introduced by Dan Linstedt and refined through nearly two decades of iterative development, is a data warehousing methodology engineered for scalability, schema agility, and comprehensive historical traceability (Linstedt & Olschimke, 2015). Unlike the Inmon and Kimball approaches, which prioritise either enterprise normalisation or analytical usability, Data Vault is distinguished by its architectural commitment to absorbing change without schema redesign and to preserving a complete, immutable audit record of all integrated data.

The methodology's defining structural innovation is the decomposition of data into three distinct component types: Hubs, which anchor stable business identities; Links, which capture relationships between those identities; and Satellites, which store descriptive attributes and their historical evolution. By separating these concerns, Data Vault isolates the volatile elements of a data landscape (i.e. attribute values and contextual metadata) from the stable elements (i.e. business keys and relational associations) enabling the architecture to evolve incrementally by extension rather than redesign (Linstedt & Olschimke, 2015).

### *3.3.2 Architectural Design and Core Components*

The Data Vault architecture is organised into three principal tiers. The Staging and Preparation layer receives raw data extracted from source systems and performs hash key generation and basic conforming without applying business transformation logic. The Raw Data Vault constitutes the core integration layer, comprising Hubs, Links, and Satellites loaded with minimal transformation, preserving a complete and auditable record of all ingested data. An optional Business Vault layer may apply business rules and derived calculations to vault structures without modifying the underlying raw data. The Data Mart layer, analogous to the presentation layer in the Kimball methodology, provides dimensional and query-optimised structures derived from vault data for end-user analytical consumption.

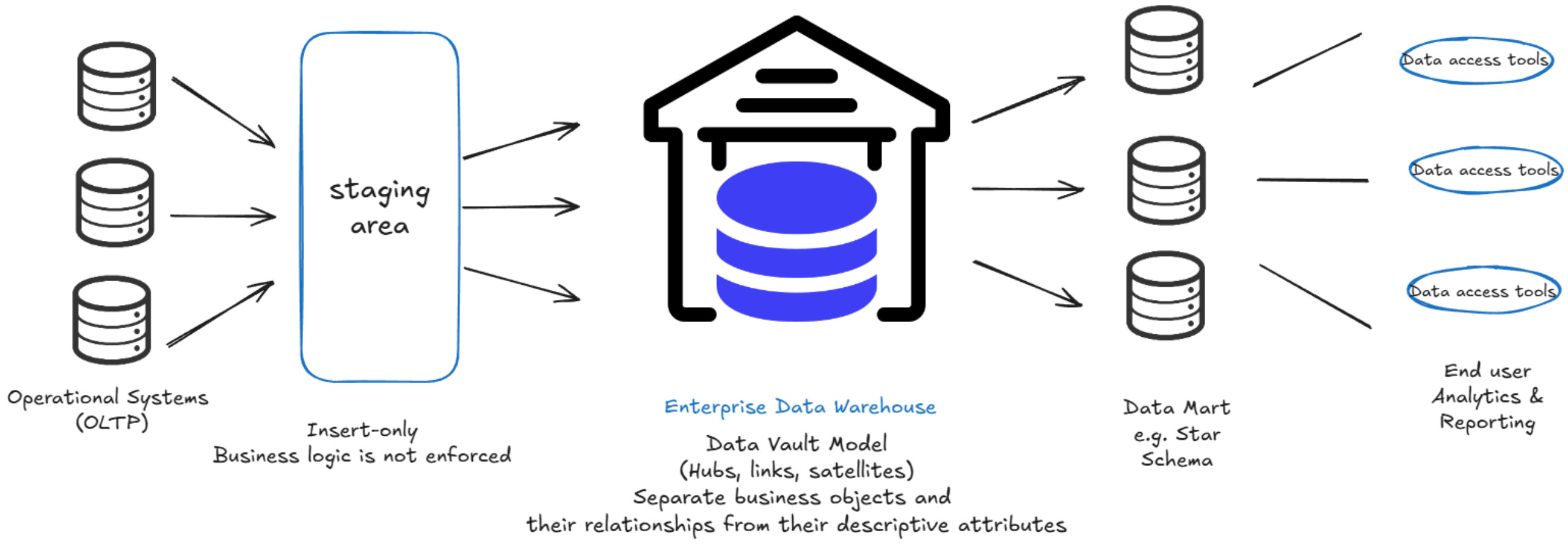

*Figure 5. Data Vault — Three-tier architecture: Staging, Data Vault, and Data Marts*

The three core structural components of the Data Vault merit detailed examination. Hubs represent the fundamental business entities of the organisation (e.g., customers, products, accounts, employees) and contain only the business key, a generated hash key for performance, a load timestamp, and a record source identifier (Figure 6). The deliberate exclusion of descriptive attributes from Hubs reflects the observation that business keys tend to remain stable over time even as contextual information evolves; by isolating keys, Hubs provide durable integration anchors that rarely require structural modification.

Links capture relationships between business entities by connecting two or more Hubs (Figure 6). A Link table records the hash keys of the associated Hubs, its own hash identifier, and standard metadata. Links contain no descriptive information and assert only that a relationship exists. This abstraction allows complex multi-entity relationships to be represented without altering existing Hub or Satellite structures, and permits the same relationship to be loaded from multiple source systems without conflict.

Satellites store the descriptive attributes and historical context associated with Hubs or Links (Figure 6). Rather than updating existing records when attribute values change, Satellites insert new rows with effective timestamps, preserving a continuous timeline of all historical states. This insert-only historization model is the mechanism through which Data Vault achieves its comprehensive audit capability: every attribute value that has ever been recorded is permanently accessible, enabling point-in-time reconstruction of the data landscape at any historical moment (Linstedt & Olschimke, 2015).

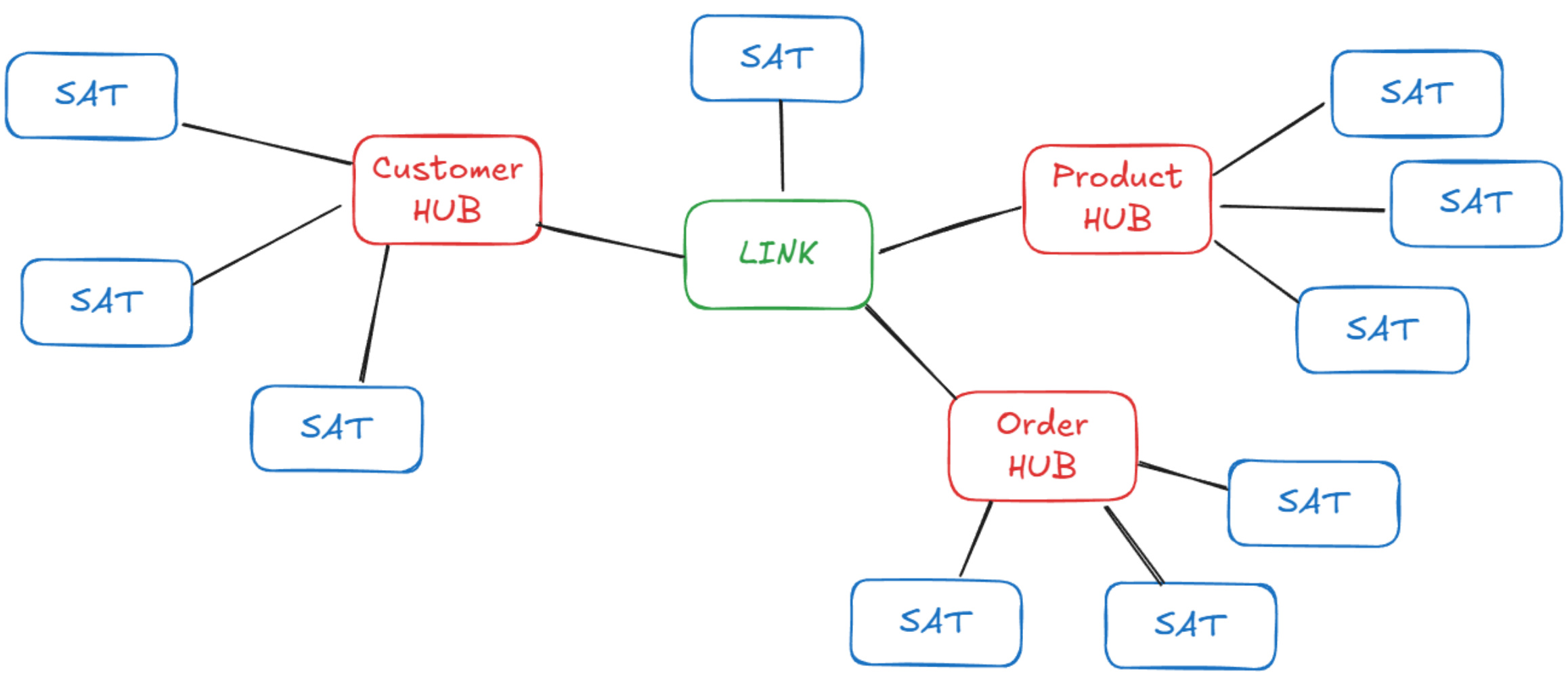

*Figure 6. Hub-Link-Satellite relationships in the Raw Data Vault*

### *3.3.3 Advantages and Limitations*

The Data Vault methodology's principal strengths are its schema agility and its auditability. Because structural changes in source systems propagate primarily to Satellite tables, with Hub and Link structures remaining intact, the architecture absorbs change by addition rather than redesign. This property makes Data Vault particularly well-suited to organisations that are subject to frequent source system evolution, merger and acquisition activity, or expanding regulatory requirements. Table 4 summarises the methodology's key advantages and limitations.

| Advantages | Limitations |
|---|---|
| • **Extreme scalability: the hub-link-satellite model extends by addition, not redesign.**<br>• **Superior schema agility: source system changes affect only peripheral satellites, leaving the core intact.**<br>• **Complete and immutable audit trail: all changes are preserved as new rows, never overwritten.**<br>• **Strong governance and lineage: every record is traceable to its source system and load timestamp.** | • High implementation complexity: requires automation tooling and specialist expertise.<br>• Data Vault is unsuitable for direct analytical querying; a downstream mart layer is always required.<br>• Higher storage consumption due to full historisation of all attribute changes.<br>• Longer path to first analytics delivery compared to a targeted Kimball mart implementation. |

*Table 4. Advantages and Limitations of the Data Vault Approach*

The methodology's primary limitation is its complexity. The hub-link-satellite model is non-trivial to implement effectively and typically requires dedicated automation tooling and practitioners with specialised expertise. Furthermore, the Data Vault is not suitable for direct analytical querying; a downstream mart layer is invariably required to serve end-user analytical needs, introducing an additional architectural tier and extending the path to first analytics delivery relative to a targeted Kimball implementation.

### *3.3.4 Illustrative Application*

To ground the Data Vault structure in a concrete example, consider an order processing system. A Customer Hub anchors the business identity of each customer, containing only a unique business key, a generated hash identifier, load metadata, and record source. A CustomerDetails Satellite, attached to the Customer Hub, stores descriptive attributes such as preferred contact method, membership tier, and geographic region. Rather than updating records when these attributes change, the Satellite inserts a new row with an effective timestamp, preserving the full historical trajectory of each customer's attributes. An OrderLine Link captures the relationship between the Customer Hub and a Product Hub, recording which products were purchased by which customers across which orders. The resulting structure is fully auditable, extensible by the addition of new Satellites or Links, and resilient to structural changes in the underlying source systems.

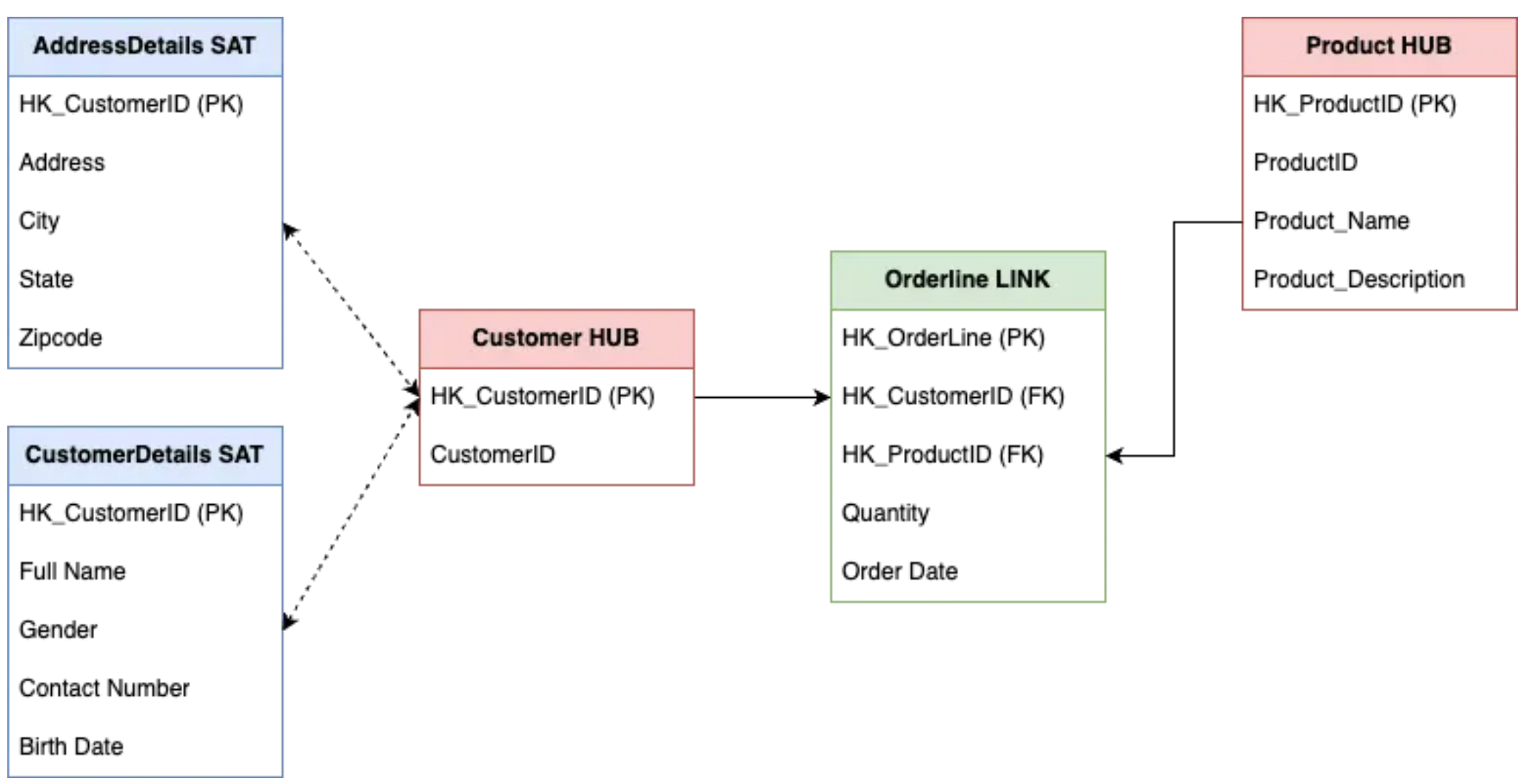

*Figure 7. Data Vault example — Customer Hub, OrderLine Link, and CustomerDetails Satellite*

# 4. COMPARATIVE BENCHMARK ANALYSIS

Having examined each methodology individually, this section presents a consolidated comparative assessment across twelve evaluative criteria: design philosophy, architectural pattern, modelling technique, time-to-value, query performance, flexibility to change, scalability, audit and historical tracking, ETL complexity, user involvement, governance maturity, and ideal use case. The objective is to provide practitioners and researchers with a concise, structured reference for methodology selection. Table 5 summarises the comparative findings.

| Criterion | Inmon (Top-Down) | Kimball (Bottom-Up) | Data Vault |
|---|---|---|---|
| **Design Philosophy** | Centralised EDW as single source of truth | Dimensional marts aligned to business processes | Separation of business keys, relationships, and context |
| **Architecture** | 3NF EDW → derived subject marts | Bus architecture of conformed dimensional marts | Staging → Data Vault (Hub/Link/Sat) → Business Marts |
| **Modelling Technique** | Relational 3NF / ER modelling | Star or snowflake schema (facts & dimensions) | Hub-Link-Satellite (HLS) modelling |
| **Time-to-Value** | Long: High upfront investment | Short: Incremental mart delivery | Medium: Architectural setup; then incremental |
| **Query Performance** | Slow on EDW; optimised via marts | High; star schema minimises joins | Slow on Data Vault; mart layer required for performance |
| **Flexibility** | Moderate: source changes require EDW extension | Low: frequent changes break ETL and dimensions | High: changes absorbed by new satellites or hubs |
| **Scalability** | High in scope; labour-intensive to extend | Scales by mart addition; consistency risk at scale | Highly extensible by design; parallel-load capable |
| **Audit & History** | Time-variant 3NF; history via record versioning | Selective; via Slowly Changing Dimensions (SCDs) | Complete and immutable; every change preserved |
| **ETL Complexity** | High initial effort; simpler per-entity mappings | Complex denormalisation and conforming per mart | Template-driven; low per-component complexity |
| **User Involvement** | IT-driven; business users consume marts | Business-driven; users co-define dimensions | Collaborative; business input on keys and mart design |
| **Governance Maturity** | High; centralised standards and lineage | Medium; depends on dimensional standards discipline | High; lineage, audit, and provenance by design |
| **Ideal Use Case** | Large regulated enterprises needing holistic EDW | SMEs or departments with stable, defined requirements | Dynamic enterprises, compliance-heavy industries |

*Table 5. Comparative Benchmark of Inmon, Kimball, and Data Vault Methodologies*

The benchmark reveals that the three methodologies occupy distinct positions on several key axes of trade-off. On the governance-agility spectrum, Inmon and Data Vault both achieve high governance maturity, but through fundamentally different mechanisms: Inmon through upfront normalisation and centralised definitional authority, Data Vault through immutable historisation and structural decomposition. Kimball's governance maturity is contingent on the discipline with which conformed dimensions are managed across marts. On the performance-flexibility axis, Kimball excels in query performance through star schema design but sacrifices flexibility; Data Vault achieves the inverse, prioritising agility at the cost of requiring a separate analytical layer. The Inmon approach occupies a middle position on both dimensions, delivering acceptable performance through derived marts while maintaining moderate schema flexibility.

These trade-offs suggest that methodology selection should be guided by an organisation's most binding constraints rather than by a single dimension of evaluation. Organisations for whom regulatory compliance and audit completeness are paramount will find Data Vault's immutable historisation architecturally compelling. Those facing well-defined analytical requirements with limited timelines will benefit from Kimball's rapid mart delivery model. Enterprises requiring a holistic, organisation-wide integration layer as a long-term strategic asset are best served by the Inmon paradigm. It is also notable that hybrid architectures, drawing on elements of multiple methodologies, are increasingly common in practice, particularly the use of a Data Vault integration layer with Kimball-style presentation marts (Linstedt & Olschimke, 2015; Kimball & Ross, 2013).

## 5. CONCLUSION

This paper has presented a technically grounded comparative analysis of the three dominant enterprise data warehousing methodologies: Inmon, Kimball, and Data Vault. The analysis was prefaced by a review of the foundational technical concepts, including OLTP and OLAP processing paradigms, relational normalisation theory, and entity-relationship and dimensional modelling techniques—that constitute the shared substrate upon which each methodology is constructed.

The comparative analysis demonstrates that each methodology embodies a coherent but distinct theory of enterprise data management. The Inmon approach delivers enterprise-wide definitional authority and long-term governance at the cost of extended time-to-value and reduced responsiveness to change. The Kimball approach accelerates analytical delivery and maximises business alignment at the cost of architectural flexibility and governance scalability. Data Vault provides the highest degree of schema agility and auditability but demands greater implementation complexity and an additional tier of transformation before analytical queries can be executed.

The principal finding of this analysis is that no single methodology is universally optimal. Effective methodology selection requires a systematic assessment of organisational context: the scale and diversity of the data landscape, the stability or volatility of source system structures, the regulatory and compliance environment, the analytical maturity of business users, and the organisation's tolerance for upfront architectural investment versus iterative delivery. In many enterprise contexts, hybrid architectures that combine elements of multiple methodologies—most commonly a Data Vault or Inmon-based integration layer with Kimball-style analytical marts—may offer the most pragmatic path to balancing governance rigour with analytical responsiveness.

Future research directions include empirical benchmarking of real-world implementations across organisations of varying scale and industry sector; the systematic evaluation of hybrid methodology architectures; and the assessment of how contemporary technologies, including cloud-native data platforms, in-memory processing engines, and automated ETL frameworks, alter the relative trade-offs examined in this analysis. As the data warehousing landscape continues to evolve, the conceptual frameworks and comparative dimensions established in this paper provide a durable foundation for ongoing methodological inquiry.

# REFERENCES

Chaudhuri, S., & Dayal, U. (1997). An overview of data warehousing and OLAP technology. ACM SIGMOD Record, 26(1), 65–74. https://doi.org/10.1145/248603.248616

Chen, P. P.-S. (1976). The entity-relationship model: Toward a unified view of data. ACM Transactions on Database Systems, 1(1), 9–36. https://doi.org/10.1145/320434.320440

Codd, E. F. (1970). A relational model of data for large shared data banks. Communications of the ACM, 13(6), 377–387. https://doi.org/10.1145/362384.362685

Codd, E. F. (1972). Further normalization of the data base relational model. In R. Rustin (Ed.), Data Base Systems (pp. 33–64). Prentice-Hall.

Arab, I.; Falah, B.; Magel, K. SCMS: Tool for Assessing a Novel Taxonomy of Complexity Metrics for any Java Project at the Class and Method Levels based on Statement Level Metrics. Adv. Sci. Technol. Eng. Syst. J. 2019, 4, 220–228. https://doi.org/10.25046/aj040629

Arab, I., Magel, K., & Akour, M. (2025). Evaluating the Predictive Power of Software Metrics for Fault Localization. Computers, 14(6), 222. https://doi.org/10.3390/computers14060222

Ravi, V. K., & Cheruku, S. R. (2024). AI and machine learning in predictive data architecture. *International Research Journal of Modernization in Engineering Technology and Science*.

Inmon, W. H. (2005). Building the data warehouse (4th ed.). Wiley.

Inmon, W. H. (2016). Data lake architecture: Designing the data lake and avoiding the garbage dump. Technics Publications.

Kimball, R., & Ross, M. (2013). The data warehouse toolkit: The definitive guide to dimensional modeling (3rd ed.). Wiley.

Linstedt, D., & Olschimke, M. (2015). Building a scalable data warehouse with Data Vault 2.0. Morgan Kaufmann.

Arab, I.; Bourhnane, S.; Kafou, F. Unifying modeling language-merise integration approach for software design. Int. J. Adv. Comput. Sci. Appl. 2018, 9, 4. https://dx.doi.org/10.14569/IJACSA.2018.090402

Falah, B.; Akour, M.; Arab, I.; M'hanna, Y. An attempt towards a formalizing UML class diagram semantics. In Proceedings of the New Trends in Information Technology (NTIT-2017), Amman, Jordan, 25–27 April 2017; pp. 21–27.

Arab, I.; Bourhnane, S. Reducing the cost of mutation operators through a novel taxonomy: Application on scripting languages. In Proceedings of the International Conference on Geoinformatics and Data Analysis, Prague, Czech Republic, 20–22 April 2018; pp. 47–56. https://doi.org/10.1145/3220228.3220264

Simsion, G. C. (2007). Data modeling: Theory and practice. Technics Publications.

Date, C. J. (2003). An introduction to database systems (8th ed.). Addison-Wesley.